# Quantum Properties of Cavity Cerenkov Radiation

Ju Gao[*] and Fang Shen
*Electrical and Computer Engineering Department, University of Illinois, Urbana, IL 61801*
(Dated: December 28, 2005)

Cerenkov radiation from cavities have been analyzed by quantum electrodynamic theory. Analytical expressions of some basic radiation properties including Einstein's $A$ and $B$ coefficients are derived and shown to be directly modified by the cavities. Coherent and incoherent radiations are analyzed with the aim of generating THz radiation from the devices.

PACS numbers: 3.50, 32.80, 42.50

## I. INTRODUCTION

A free traveling electron emits photons spontaneously when its speed $u$ is greater than the phase velocity $v_p$ of the photon it emits. Such process can occur in a dielectric medium [1, 2] and is known as Cerenkov radiation (CR). Since its discovery, Cerenkov radiation has played an important role in high energy physics for detecting particles [3]. The broad spectrum of CR has also stimulated thoughts [4] of using it as a radiation source, particularly in a frequency range difficult of assess by other means. High frequency microwave radiations in the hundreds of GHz range have been generated in dielectrically loaded CR devices [5, 6] where a vacuum tunnel is typically used inside the dielectric for the electrons to travel and a metal cavity is also used to confine the radiation field. Here the topic is revisited because there have been much interests recently in generating practical THz radiation, which ranges between 300 GHz to 30 THz in frequency. What makes THz radiation particularly interesting is the natural match of the frequency band to the molecular vibrational and rotational energy bands, leading to potential applications in chemistry, biology and astronomy, etc. The simplicity of the radiation scheme, mature technology of fabricating dielectric structures and the possibility of integrating field emission electron sources [7] present CR device as yet another alternative to the pursuit of submillimeter or THz radiation, in parallel to synchrotrons [8], free electron lasers [9], optically-pumped molecular lasers [10], quantum-cascade lasers [11], and femtosecond laser-pumped photoconductive antennas [12].

Analysis of the CR devices [13–19] are mostly treated by classical electrodynamics, where the electron motion is governed by the Newton-Lorentz equation and the radiation as a result of the moving electron is ruled by the Maxwell equations. The treatment is justified for the lower frequency range. In the higher frequency range, from infrared (IR) to ultraviolet (UV), quantum theory gives a more accurate description of quantum electronics devices such as lasers. Since THz radiation is an extension of IR, quantum mechanical treatment for the radiation is adequate and even required. In this paper, we will analyze the basic radiation properties of the CR devices. For example, we will calculate the Einstein's $A$ and $B$ coefficients and show how the cavity affects their value. We will also utilize the numbers to analyze the incoherent and coherent radiations from the device.

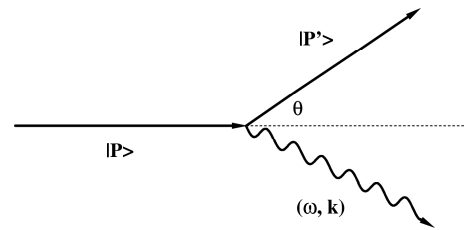

FIG. 1: A Cerenkov photon of frequency $\omega$ and wave vector $\mathbf{k}$ is generated by the transition of free electron from $|\mathbf{P}>$ state to $|\mathbf{P}'>$ state, where $\mathbf{P}$ and $\mathbf{P}'$ represent the electron momentum before and after the radiation.

## II. EINSTEIN'S $A$ COEFFICIENT FOR CR IN A UNIFORM DIELECTRIC MEDIUM

We start with deriving the Einstein's $A$ coefficient in a uniform dielectric medium. We will show that the quantum treatment results in some different conclusions but asymptotically approaches to the classical CR power expression.

The basic CR has the electron travel inside a uniform dielectric medium. In the quantum picture, a photon is generated by the electron with energy and momentum conservation,

$$(E_{P'}, \mathbf{P}) = (E_P, \mathbf{P}) + (\hbar\omega, \hbar\mathbf{k}) \tag{1}$$

where $(E_{P'}, \mathbf{P})$ and $(E_P, \mathbf{P})$ are the 4-momenta of the electron before and after the process, respectively, with $E_P = \sqrt{\mathbf{P}^2 + m^2}$ and $E_{P'} = \sqrt{\mathbf{P}'^2 + m^2}$. $\omega$ and $\mathbf{k}$ are the photon frequency and wave vector, respectively. The process can be illustrated by a diagram as shown in Fig. 1. Notice a photon can not be generated in the vacuum according to Eq. 1.

---

[*]Electronic address: jugao@uiuc.edu

The ability to radiate is measured by the Einstein's $A$ coefficient, a concept that was introduced by Einstein even before the full repertoire of quantum electrodynamics (QED) was developed. $A$ is equivalent to the transition probability rate,

$$\begin{aligned} A &= 2\pi |H_{int}|^2 \delta(E_{P'}/\hbar + \omega, E_P/\hbar) \\ &= 2\pi (\frac{1}{\hbar})^2 |<1|<\mathbf{P'}|c\alpha \cdot \mathbf{A}(\mathbf{k}\cdot\mathbf{r})|\mathbf{P}>|0>|^2 \\ &\quad \delta(E_{P'}/\hbar + \omega, E_P) \end{aligned} \quad (2)$$

where $H_{int}$ is the transition matrix and $\delta$ function enforces the resonant condition. $|\mathbf{P}> = \frac{1}{\sqrt{V}} e^{\mathbf{P}/\hbar\cdot\mathbf{r}} u(\mathbf{P})$ and $|\mathbf{P'}> = \frac{1}{\sqrt{V}} e^{\mathbf{P'}/\hbar\cdot\mathbf{r}} u(\mathbf{P'})$ are the electron wavefunctions before and after the radiation. $V$ is the normalization volume and $u(\mathbf{P})$ and $u'(\mathbf{P})$ are the Dirac spinors. A photon is generated, $|0> \to |1>$, by the interaction $c\alpha \cdot \mathbf{A}(\mathbf{k}\cdot\mathbf{r})$ in which dipole approximation is not used. $\alpha$ is a Dirac matrix and $\mathbf{A}(\mathbf{k}\cdot\mathbf{r})$ is the quantized radiation field given by

$$\mathbf{A}(\mathbf{k}\cdot\mathbf{r}) = g(\epsilon\,\hat{a}\,e^{i\mathbf{k}\cdot\mathbf{r}} + \epsilon^*\hat{a}^\dagger e^{-i\mathbf{k}\cdot\mathbf{r}}). \quad (3)$$

where $\hat{a}$ and $\hat{a}^\dagger$ are the creation and annihilation photon operators, respectively. $e^{i\mathbf{k}\cdot\mathbf{r}}$ represents the plane wave photon field with, however, the dispersion relationship is $k = n\frac{\omega}{c}$ where $n$ is the index of refraction of the medium. $g = \sqrt{\frac{\hbar}{2\epsilon V_\gamma \omega}}$ is the field normalization constant where $V_\gamma$ is the normalization volume of the field.

Equation 3 is readily carried out after integrating over all $\mathbf{k}$s and $\mathbf{P'}$s, where the radiation angle is derived to be frequency dependent $\cos\theta = \frac{1}{\beta n} + \frac{\hbar k}{2P}(1 - \frac{1}{n^2})$. Uniform dispersion is assumed for the medium. The analytical expression of the Einstein's $A$ coefficient for the CR in the uniform medium is obtained,

$$A = \alpha\beta\frac{E_P}{\hbar} F(\beta, n) \quad (4)$$

where $\alpha = \frac{e^2}{4\pi\epsilon_0 \hbar c}$ is the fine structure constant and $\beta = u/c$, where $u$ is the electron speed. $F(\beta, n)$ is an explicit function of $\beta$ and $n$,

$$F(\beta, n) = (\frac{7}{4\beta^2} + \frac{1}{4\beta^2 n^2} - 2) + (\frac{3}{8\beta^2} + \frac{1}{8\beta^2 n^2} - \frac{1}{2})\eta_m + (\frac{7}{4\beta^2} + \frac{5}{4\beta^2 n^2} - 3)\ln(1-\eta_m)/\eta_m \quad (5)$$

where $\eta_m \equiv \frac{\hbar\omega_m}{E_P}$ and $\omega_m$ is the maximal radiation frequency,

$$\hbar\omega_m = 2\frac{\beta n - 1}{n^2 - 1} E_P < \frac{2}{n+1} E_P < E_P. \quad (6)$$

Equation 6 shows that the electron can not convert its entire energy to a radiating photon, in contrast to the infinite maximum frequency claimed by the classical theory. If we allow $\omega_m$ to be infinite, however, we recover the expression for the radiation power identical to the classical results for $N_e$ electrons,

$$P = A\hbar\omega N_e = \frac{e^2 N_e}{4\pi\epsilon_0 c}\beta \int_0^\infty \omega d\omega(1 - \cos^2\theta). \quad (7)$$

To appreciate the value of the $A$ coefficient, let the material be quartz so $n = \sqrt{3.78} = 1.944$ and $\beta = .634 > 1/n$. The maximal photon frequency is $\omega_m = 1.7 \times 10^{20}\,\mathrm{rad\,s^{-1}}$ according to Eq. 6. Figure 2 shows $A$ value as a function of the cutoff frequency up to $\omega_m$. In reality, the medium becomes absorptive at such high frequency, thus cutoff frequency of CR is much smaller. Suppose the cutoff wavelength is 100nm, we have $A = 1.55 \times 10^{13}\mathrm{s^{-1}}$, which is quite large compared with an atomic transition. Typically $A$ is for line transitions, However, the $A$ value here represents the total radiation within a large band.

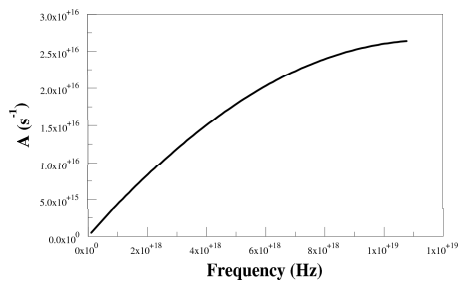

FIG. 2: Einstein's $A$ coefficient values as a function of the cutoff wavelength for $n = 1.944$ and $\beta = .634$.

### III. EINSTEIN'S $A$ AND $B$ COEFFICIENTS FOR CAVITY CERENKOV RADIATION

It is desirable for many applications to have the radiation energy concentrated in a narrow and discrete band, which requires discrete energy levels of the radiation system. The free electron does not possess discrete energy levels for discrete radiations, but the alternative is to force discrete fields by a cavity so that the electron can only loose its energy to those fields. A cavity CR device has thus been formed by enclosing the dielectric with a conducting material [6, 20] as shown in Fig. 3, where a vacuum tunnel is built for the electrons to travel. We will study the structure illustrated by Fig. 3 as a basic cavity CR device.

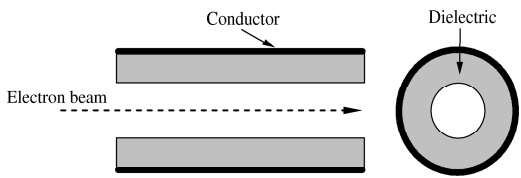

FIG. 3: A typical dielectric lined cavity CR device illustrates a conducting tube with radius b and a vacuum tunnel with radius a.

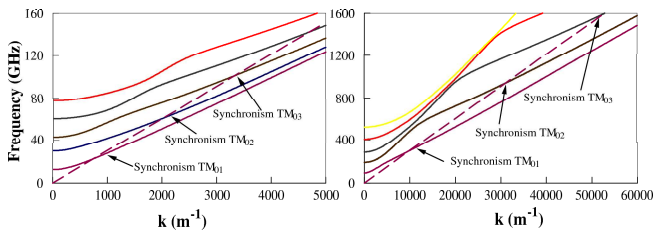

FIG. 4: Dispersion curves for the different modes in two different cavity CR devices:(a) $a = 3.$mm, $b = 6.$mm ; (b) $a = 1.0$mm, $b = 1.2$mm. Both have $\epsilon = 3.78\epsilon_0$ and $\beta = .634$. The dotted line represents the linear dispersion that gives phase velocity of $\beta$.

The fields that can survive inside the cavity is called eigen mode fields that have a special dispersion relationship derived to be

$$\frac{I_1(Xa)}{I_0(Xa)X} = -\frac{\epsilon[Y_1(Ya)J_0(Yb) - J_1(Ya)Y_0(Yb)]}{\epsilon_0 Y[J_0(Ya)Y_0(Yb) - Y_0(Ya)J_0(Yb)]} \quad (8)$$

where $J_n$ is the $n$th-order Bessel function and $I_n$ and $Y_n$ are the $n$th-order modified Bessel function of the first and second kind. $X = k^2 - (\frac{\omega}{c})^2$ and $Y = \frac{\epsilon}{\epsilon_0}(\frac{\omega}{c})^2 - k^2$ are separation constants. Figure 4 plots the dispersion relationships expressed by Eq. 8 for two different cavity structures. The first cavity is used in the experiments [6, 20], and the second cavity is smaller designed for higher frequencies. Only TM modes are considered because they dominate the interaction $\mathbf{P} \cdot \mathbf{A}$.

The mode characteristic alone does not fixate discrete frequencies. Additional relation comes from energy-momentum conservation conditions between the photon and the electron described by Eq. 1. By noticing $\mathbf{k} \parallel \mathbf{P}$ and $\mathbf{P'} \parallel \mathbf{P}$, we have

$$\begin{aligned} v_p \equiv \frac{\omega}{k} &= \frac{\beta c}{1 + \frac{\eta}{2}[(\frac{\omega}{k})^2 - 1]} \\ &\approx u - u\frac{\eta}{2}[(\frac{1}{\beta})^2 - 1] \end{aligned} \quad (9)$$

where $\eta = \hbar\omega/E_P$. Eq. 9 is combined with Eq. 8 to give the actual, discrete radiation frequencies, or the synchronism frequencies because the electron velocity almost matches the field phase velocity. These points are shown as the interception points in Fig. 4 between the dispersion curve and the $\omega/k = \beta$ line. It is shown that the radiation frequencies are already in the THz range for the smaller cavity.

The cavity not only selects certain field modes but causes the field distribution to deviate from the plane wave. Because the transition rate depends on the overlap between the field function and the electron wavefunction (Eq. 2), the modified field distribution can dramatically change the radiation properties. This effect has been studied extensively as a subject of cavity QED. For example, it has been observed that the spontaneous emission rate is modified from that in vacuum [21]. The novelty here is that the radiators are the free electrons instead of atoms, molecules and even nuclei. From practical point of view, using free electrons as the radiators enable changing the energy system conveniently with external fields. Clearly the dielectric medium is a necessity for the free electron to radiate in the cavity.

The expression of the quantized radiation field inside the cavity is found to be,

$$\mathbf{A}(\mathbf{r}) = \mathbf{z} g'(\hat{a} e^{ikz} + \hat{a}^\dagger e^{-ikz}) \begin{cases} I_0(X\rho); & \rho \leq a \\ \frac{\epsilon_0}{\epsilon}\frac{Y}{X}[\frac{I_1(Xa)J_0(Yb)Y_0(Y\rho) - I_1(Xa)Y_0(Yb)J_0(Y\rho)}{Y_1(Ya)J_0(Yb) - J_1(Ya)Y_0(Yb)}]. & a < \rho \ll b \end{cases} \quad (10)$$

where the normalization coefficient $g'$ becomes,

$$g' = \sqrt{\frac{\hbar}{2\epsilon L\pi b^2 \omega}} f$$
$$f \equiv \sqrt{\frac{2\epsilon_0 \pi b^2}{\mu_0 \int_0^a H_I^2 2\pi\rho d\rho + \epsilon_0 \int_0^a E_I^2 2\pi\rho d\rho + \mu \int_a^b H_{II}^2 2\pi\rho d\rho + \epsilon \int_a^b E_{II}^2 2\pi\rho d\rho}}, \quad (11)$$

where $L$ is the cavity length and $H_I, E_I$ and $H_{II}, E_{II}$ are the magnetic and electric fields in the tunnel ($\rho \leq a$) and dielectric ($a < \rho \ll b$) regions, respectively, given explicitly by

$$\begin{aligned} E_I &= i\, I_0(Xr) \\ H_I &= \frac{\omega\epsilon_0}{X} I_1(Xr) \\ E_{II} &= i\frac{\epsilon_0}{\epsilon}\frac{Y}{X}\frac{I_1(Xa)J_0(Yb)Y_0(Yr) - I_1(Xa)Y_0(Yb)J_0(Yr)}{Y_1(Ya)J_0(Yb) - J_1(Ya)Y_0(Yb)} \\ H_{II} &= \frac{\omega\epsilon_0}{X}\frac{I_1(Xa)J_0(Yb)Y_1(Yr) - I_1(Xa)Y_0(Yb)J_1(Yr)}{Y_1(Ya)J_0(Yb) - J_1(Ya)Y_0(Yb)} \end{aligned} \quad (12)$$

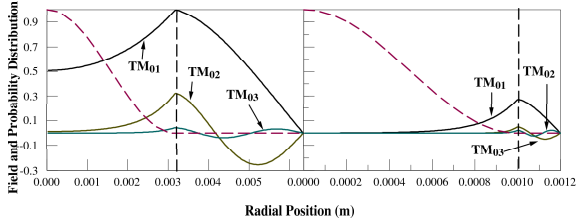

FIG. 5: The **A** field distributions (solid lines) are plotted along with the electron probability distributions (dashed line)inside the two cavities (a) $a = 3.175$mm, $b = 6.35$mm ; (b) $a = 1.0$mm, $b = 1.2$mm.

The normalized fields are plotted in Fig. 5 together with the electron wavefunctions that is also confined by the cavity as well. The difference is that the dielectric wall is the only boundary for the electrons assuming there is no tunnelling into the dielectric medium. The lowest order electron wavefunction is then,

$$\begin{aligned} |\mathbf{P}> &= \frac{e^{iPz/\hbar}u(P)}{\sqrt{L}}\frac{J_0(2.405\rho/a)}{a\sqrt{\pi}|J_1(2.405)|}; (\rho < a) \\ &= 0, (\rho > a). \end{aligned} \quad (13)$$

The corresponding electron probability distribution is plotted in Fig. 5. A narrower electron distribution can be described by the superposition of a few higher order electron wavefunctions. The electron energy and momentum are practically unchanged by the cavity because the dimension of the radial confinement is too large so that $\Delta P_\rho \sim \hbar/\Delta\rho \ll P$.

The $A$ coefficient is then calculated by using the new wavefunction (Eq. 13) and field (Eq.10), and the analytical expression is found again,

$$A_{cav} = \alpha\beta^2(1+\eta)f^2 F^2 \frac{c^2}{(2\pi)^2 b^2 \omega_0}\frac{2c}{v_g} \quad (14)$$

where $\omega_0$ is the synchronism frequency and $F$ is the filling factor that measures the overlap between the field and the electron wavefunction,

$$F = \int_0^a \frac{1}{a^2\pi J_1^2(2.405)} J_0^2(2.405\rho/a) I_0(X\rho) 2\pi\rho d\rho, \quad (15)$$

and $v_g = \frac{d\omega}{dk}|_{\omega_0}$ is the group velocity at the synchronism frequency.

The $A_{cav}$ is shown to be determined by the cavity geometry and the dielectric material. That means in practice its value can be engineered through the cavity design. Again the concept here draws analogy to that in cavity QED where the radiation phenomena are modified by the cavities.

The Einstein's $B$ coefficient can also be derived from QED by calculating the transition rate from $N_\gamma$ existing photons to $N_\gamma + 1$ photons,

$$\begin{aligned} &2\pi(\frac{1}{\hbar})^2 |<N_\gamma+1|<\mathbf{P}'|c\alpha\cdot\mathbf{A}(\mathbf{k}\cdot\mathbf{r})|\mathbf{P}>|N_\gamma>|^2 \\ &\delta(E_{P'}/\hbar + \omega, E_P) \\ &= B_{cav,emi}\,\rho_\gamma, \end{aligned} \quad (16)$$

where $N_\gamma$ is the photon number in the mode and $\rho_\gamma$ is the corresponding photon density. Equation 16 gives the Einstein's $B$ coefficient for emission. The $B$ coefficient for absorption can also be found by assuming photons go from $N_\gamma$ to $N_\gamma - 1$. Thus we have

$$B_{cav,emi} = N_\gamma \alpha \beta^2 (1+\eta) f^2 F^2 c^3/(b^2 \omega_0)/I$$
$$B_{cav,abs} = N_\gamma \alpha \beta^2 (1-\eta) f^2 F^2 c^3/(b^2 \omega_0)/I \quad (17)$$

Notice that the two $B$s are not exactly the same and the emission coefficient is only slightly bigger than the absorption one, which suggests that the device can be used as an amplifier where the stimulated emission should exceed the simulated absorption.

## IV. NUMERICAL EXAMPLES AND DISCUSSION

Numerical values of the cavity coefficient $A_{cav}$ can be calculated according to Eq. 14. Table IV lists the $A_{cav}$ values calculated for the two cavity designs. The electron energy is chosen to be 100Kev, a value used in the experiments. The result shows that the higher order modes own the smaller $A$ values, but these $A$ values are generally higher than the typical $A$ values for atoms or molecules at the same frequency range [22]. There are two main reasons for that fact: first the cavity helps to confine the radiation field so that the overlap between the field and the electron wavefunction is enhanced; second the electron speed or momentum is much larger than the electron momentum inside an atom. The latter contributes to the $A$ coefficient due to the fact that the interaction is proportional to $\mathbf{P} \cdot \mathbf{A}$, where $\mathbf{P}$ is the electron momentum (which is expressed by the operator $c\alpha$ in Eq. 2).

As shown by Eq. 14, $A_{cav}$ is proportional to $\omega^{-1}$, which indicates that scaling up the frequency of the cavity microwave devices results in weaker radiation. In the same time, notice that the atomic $A$ coefficient in the open space is proportional to $\omega^3$, which also shows the unfavorable tendency of scaling down the frequencies of the visible or IR devices. This is one of the contributing factors of the difficulty in generating THz radiation that is falling in the gap between the microwave and IR radiation.

Now that we have calculated the $A$ and $B$ coefficients, we are ready to discuss coherent radiations from quantum mechanical perspective. Let us assume the electrons are mono-energetic, neglecting the spread caused by the thermal and space charge effects. In an analogy to a laser, it appears that the population inversion is automatically achieved because all electrons occupy the same state $|\mathbf{P}>$ and the lower states $|\mathbf{P}'>$ are empty. However in this case the higher states are empty too, and the electron can make a transition to the higher states by absorbing a photon. Thus the amplified stimulated emission is determined by the competition between the stimulated emission and absorption processes. In fact, the amplification is proportional to $(B_{emi} - B_{abs})I\ N_e$, where $I$ is the intensity itself. From Eqs. 17, we find $(B_{emi} - B_{abs}) \propto \eta \equiv \hbar/E_p << 1$. The minute gain in the amplified stimulated emission is confirmed by the experiments [6, 20] when the cavity CR device is used as an amplifier. The conclusion is that the amplified stimulated emission, or lasing if light is interpreted in a broader sense, can not be the main responsibility of coherent radiation from this type of devices.

TABLE I: $A$ coefficient at different frequency for Walsh's cavity

| Mode | Frequency (GHz) | $A$ (s$^{-1}$) |
|---|---|---|
| TM$_{01}$ | 21.9 | $1.07 \times 10^7$ |
| TM$_{02}$ | 60.7 | $4.53 \times 10^5$ |
| TM$_{03}$ | 101.5 | $2.07 \times 10^3$ |

TABLE II: $A$ coefficient at different frequency for the designed cavity

| Mode | Frequency (GHz) | $A$ (s$^{-1}$) |
|---|---|---|
| TM$_{01}$ | 296 | $1.07 \times 10^6$ |
| TM$_{02}$ | 944 | $6.48 \times 10^2$ |
| TM$_{03}$ | 1601 | $1.88 \times 10^1$ |

We now turn our attention to another coherent radiation generation mechanism. Coherent radiation can indeed be generated by the spontaneous radiation from radiators occupying a space whose dimension is smaller than the radiation field wavelength. The phenomenon has been analyzed [23] even before the advent of laser and is known as the superradience or super radiation. In super radiation, the radiators interact with the vacuum fields of the same phase thus the amplitudes of the transition matrix elements for all radiators are added so that the power is proportion to the square of the number of radiators. The effect has been well studied for atoms which are immobile compared to the speed of light. For electron devices, the electrons need to be grouped together while travelling, and process is known as bunching. For our interest, a successfully bunched beam has the output power $P = A_{cav} \hbar \omega N_e^2$, where $N_e$ is the number of the bunched electrons. Experiments [5] have shown that many orders of magnitude higher output power can be achieved in the CR device by using a bunched electron beam. This is true for many other free electron radiation devices. As an example for THz radiation generation, assuming $A = 6.48 \times 10^2 \text{s}^{-1}$ for the TM$_{02}$ mode in the smaller device (Table II), we find that $5 \times 10^7$ bunched electrons in the cavity is needed to give 1mW power from the device. The current level for that number of electrons in a cavity of 20cm in length and $\mu = 0.635c$ electron speed is 7.5 mA, showing some feasibility for a practical device.

## V. CONCLUSION

We have made a quantum electrodynamic approach to calculate and interpret some basic radiation properties of the cavity CR radiation. Analytical expressions for the Einstein's $A$ and $B$ coefficients of the device are explicitly derived, which should facilitate the analysis of this type of devices parallel to that of quantum electronic devices. We point out that the cavity effect on the radiation properties of the CR device is of the same nature of the cavity QED, except the radiator is a free travelling electron here. This implies more quantum field effects may be explored. The justification of investigating the quantum nature of the radiation device lies in the fact that THz radiation is an immediate extension of the visible and IR radiations.

The authors would like to thank P. D. Coleman for the discussion of his work on the Cerenkov radiation.